\documentclass[letter]{aa}
\usepackage{graphicx} 
\usepackage{natbib}
\usepackage{txfonts}
\begin{document}
\headnote{Letter to the Editor}
\title{Evidence for a hot dust-free inner disk around 51 Oph
  \thanks{Based on observations collected at the European Southern
    Observatory at La Silla and Paranal, Chile (ESO Programme
    68-C-0474)}}

   \author{W.-F. Thi 
           \and
           B. van Dalen
           \and
           A. Bik
           \and
           L.~B.~F.~M. Waters 
           }
           
  \institute{Sterrenkundig Instituut Anton Pannekoek,
             University of Amsterdam, Kruislaan 403 1098 SJ Amsterdam
             the Netherlands}

   \offprints{Rens Waters}
   \date{Received .../Accepted ...}
   
   \abstract{We report on the observation of CO bandhead emission
     around \object{51 Oph} ($\Delta v=2$). A high resolving power
     ($R\simeq$10,000) spectrum was obtained with the infrared
     spectrometer ISAAC mounted on $VLT-ANTU$. Modeling of the profile
     suggests that the hot ($T_{\rm gas}=$ 2000--4000~K) and dense
     ($n_{\mathrm{H}}>10^{10}$ cm$^{-3}$) molecular material as probed
     by the CO bandhead is located in the inner AU of a Keplerian disk
     viewed almost edge-on. Combined with the observation of cooler
     gas ($T_{\rm gas}=$ 500-900~K) by ISO-SWS and the lack of cold
     material, our data suggest that the disk around \object{51 Oph}
     is essentially warm and small. We demonstrate the presence of a
     dust-free inner disk that extents from the inner truncation
     radius until the dust sublimation radius. The disk around
     \object{51 Oph} may be in a rare transition state toward a small
     debris disk object.  }

\maketitle

%

\section{Introduction}

Disks are commonly found around low- and intermediate-mass
pre-main-sequence stars and are understood as natural byproducts of
the conservation of angular momentum during star formation. The
structure of the inner few Astronomical Units (AU) of protoplanetary
disks plays an important role in the transfer of matter onto the star
and the formation of terrestrial planets.  The temperature in the
inner AU of disks can attain a few thousand Kelvin and the density can
be greater than 10$^{12}$ cm$^{-3}$. Those conditions are required for
CO bandheads ($\Delta v=2$) to emit.  Indeed, CO bandhead emission has
been detected around young low- and high-mass stars (e.g.,
\citealt{BikThi2004A&A}; and \citealt{Najita2000prpl.conf..457N} for
an overview). The inner disk is likely gas-rich but dust-free since
dust grains evaporate at $T\simeq$~1500~K.

\object{51 Oph} is a peculiar Be (B9.5IIIe) star located at 130 pc
with an observed rotation velocity $v\sin i = 267 \pm 5\ \mathrm{km
  s}^{-1}$ \citep{Dunkin1997MNRAS.290..165D}; however, the exact
spectral type is still disputed. The Spectral Energy Distribution
(SED) shows a strong near- till mid- infrared excess but contrary to
its lower-mass counterparts the Herbig~Ae stars, the flux in the
(sub)-millimeter domain is low, indicating a lack of small cold dust
grains.  \object{51 Oph} is the only Be star in a sample of 101 whose
infrared excess is dominated by warm dust and not by free-free
emission \citep{Waters1988A&A...203..348W}.  Both fits to the SED and
direct imaging of the disk around \object{51 Oph} agree on the
compactness of the disk (\citealt{Jayawardhana2001AJ....122.2047J};
\citealt{Leinert2004A&A...423..537L};
\citealt{Waters1988A&A...203..348W}). The disk probably does not
extent beyond 100~AU. Silicate emission features were first detected
by \cite{Fajardo1993ApJ...417L..33F} and further analyzed by
\cite{Bouwman2001A&A...375..950B}.  The UV and optical part of the SED
are well fitted by a Kurucz model and show no excess testifying that
the accretion of matter onto the star is low
(\citealt{Waters1988A&A...203..348W};
\citealt{Malfait1998A&A...331..211M}). However, infalling ionic and
atomic gases were observed which prompted
\cite{Grady1991ApJ...367..296G} to claim that \object{51 Oph} is in
the same evolutionary state as \object{$\beta$~Pic} (see also
\citealt{Roberge2002ApJ...568..343R}).  Contrary to
\ \object{$\beta$~Pic}, large amount of molecular gas (CO, CO$_2$,
H$_2$O and NO) has been seen around \object{51 Oph} by the
Short-Wavelength-Spectrometer on board the Infrared Space Observatory
(ISO) \citep{Ancker2001A&A...369L..17V}. A controversy remains about
the quantity of CO. Absorption lines by CO in the Far-Ultraviolet were
not detected by FUSE \citep{Roberge2002ApJ...568..343R}. Nevertheless,
all studies suggest that the circumstellar matter is most likely in
the form of a Keplerian disk rather than in a spherical shell.
Finally, the fractional excess IR luminosities $L_{\rm IR}/L_*$ of
\object{51 Oph} is 0.028, a value between the younger Herbig Ae stars
($>$0.1) and the Vega-like objects (10$^{-5}$--10$^{-3}$).  We
obtained a high resolution $K$-band spectrum of \object{51 Oph}
centered around the position of a CO bandhead emission in order to
clarify the detection of large amount of CO ($N(\mathrm{CO})= (0.1-10)
\times 10^{21}\ \mathrm{cm}^{-2}$) by ISO and to characterize the
structure of the inner dust-poor molecular disk. After describing the
observation and data reduction in Section \ref{Observations}, a fit by
a synthetic spectrum emitted by a Keplerian disk is shown in Section
\ref{Results}.  The discussion in Section \ref{Discussion} focuses on
the structure of the inner disk around \object{51 Oph}.

%

\section{Observation and Data Reduction}\label{Observations}

The $K$-band spectrum was obtained with the near-infrared facility
ISAAC mounted on the {\em Very Large Telescope} (VLT-ANTU) on March
15$^{\mathrm{th}}$ 2002 at resolving power of 10,000 (slit width of
0.3\arcsec) in Service mode.  The spectrum toward \object{51 Oph} is
part of a survey of CO bandhead in Herbig~Ae stars. The data were
reduced in a standard way using IRAF. We made use of flatfield and arc
frames taken by the ESO staff during the day.  Standard stars of
spectral type A observed at similar airmass than \object{51~Oph} were
used to correct for telluric OH emission and absorption.

   \begin{figure}
     \centering
     \resizebox{\hsize}{!}{\includegraphics[angle=90]{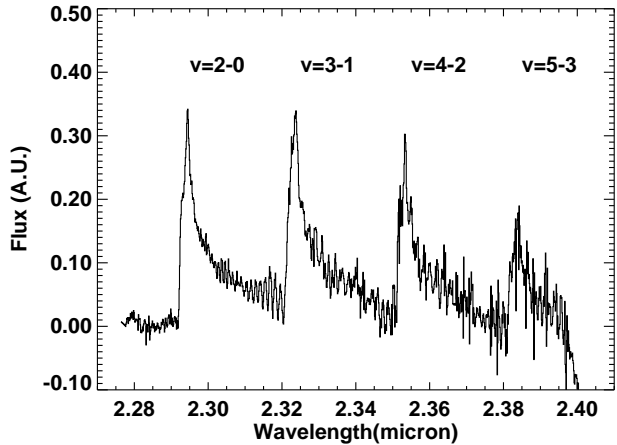}}
   \caption{Continuum subtracted CO bandheads toward \object{51 Oph} in arbitrary units. Four bandheads are detected. The spectrum of the last two bandheads is heavily blended.}
              \label{Fig_51Oph_4bandheads}%
    \end{figure}
\section{Results and Modeling}\label{Results}

The continuum subtracted and normalized spectrum is shown in
Fig.~\ref{Fig_51Oph_4bandheads}. Four bandheads are detected with high
signal-to-noise ratio (S/N$>$100), namely $\Delta v = 2 - 0$ ($\lambda
\simeq 2.2935\ \mu$m), $\Delta v = 3 - 1$ ($\lambda \simeq 2.3227\ 
\mu$m), $\Delta v = 4 - 2$ ($\lambda \simeq 2.3535\ \mu$m), $\Delta v
= 5 - 3$ ($\lambda \simeq 2.3829\ \mu$m). The first bandhead shows a
prominent blue wing. The higher bandheads are contaminated by emission
from the $P$-branch from lower bandheads.
\begin{figure}
\centering
\resizebox{\hsize}{!}{\includegraphics[]{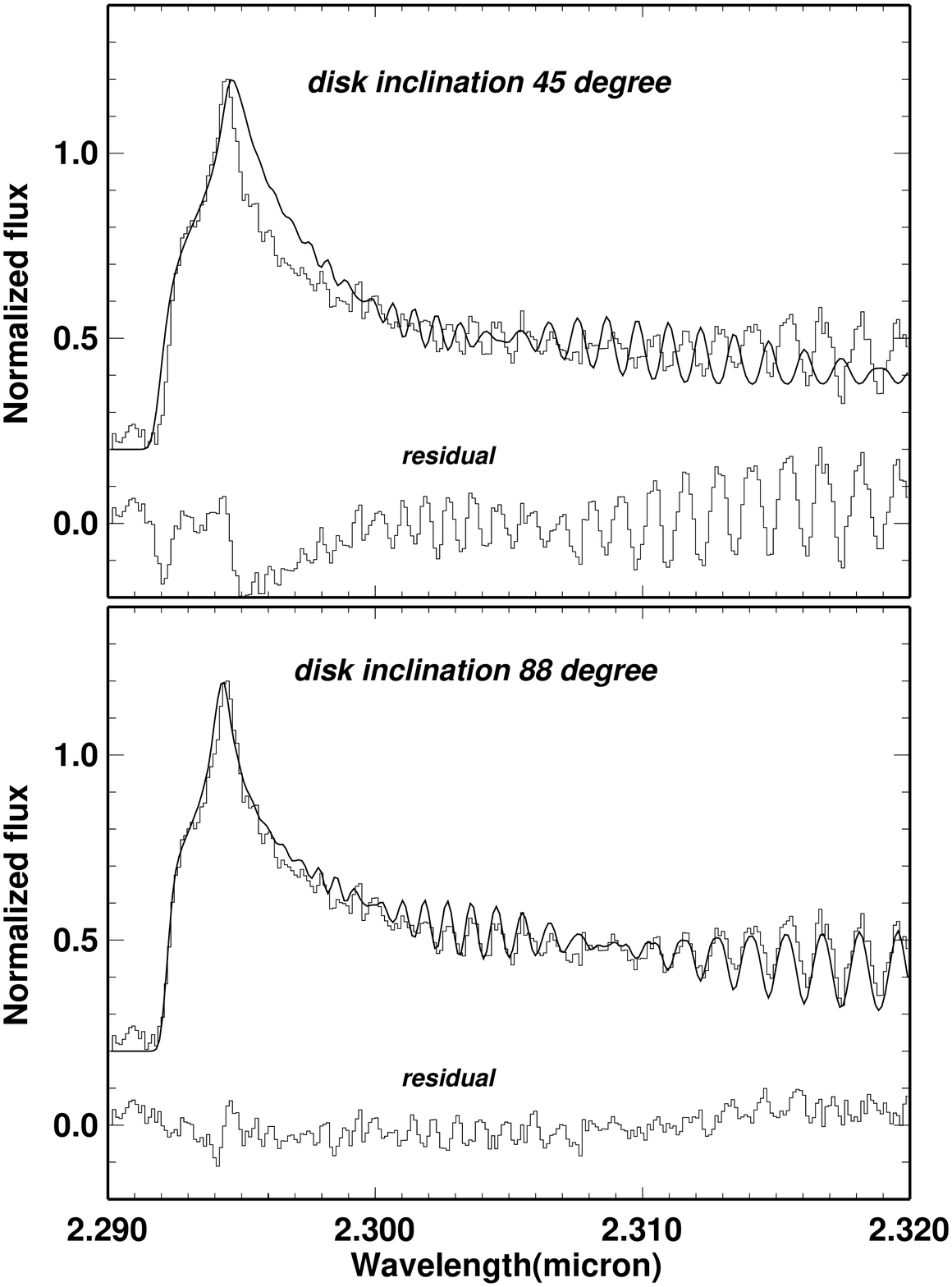}}
\caption{Fits to the CO first bandhead ($v=2-0$). In the upper panel the thin line corresponds to the normalized observed spectrum while the thick line is the best fit if the disk is viewed at an inclination of 45~$\degr$. The lower plot is the residual of the fit. The lower panel shows the best fit obtained with a disk viewed at an inclination of 88~$\degr$. The complete set of parameters is reported in Table~\ref{51Oph_table_fit}}
\label{Fig_51Oph_fit}%
\end{figure}
%
    
We decided to fit the first bandhead only since the emission from
higher bandheads is blended and contributions from different bandheads
are difficult to disentangle.  A spherically symmetric and radially
expanding wind results in flat-topped profiles for optically thin
lines \citep{Beals1931MNRAS..91..966B}.  A flat-topped profile cannot
reproduce the blue-wing seen in our spectrum; therefore our data
definitely rule out hot gas in an expanding shell. The signature of CO
bandhead emission from a narrow Keplerian rotating disk is commonly
found around young stars over a large range of masses.  Therefore, we
constructed a simple model that is detailed below. The distribution of
CO lines generated by a plane-parallel isothermal slab is convolved
with the double-peaked profile produced by a narrow ring extending
from $R_{\mathrm{min}}$ to $R_{\mathrm{max}}$.  The ring rotates at
Keplerian velocity $v_{\mathrm{Kep}}=\sqrt{GM_{*}/r}$. The line
frequencies and Einstein coefficients are provided by
\cite{Chandra1996A&AS..117..557C}. The rotational levels within each
vibrational level are assumed to be in local thermodynamic equilibrium
because the critical density for thermalization of the rotational
levels is relatively low ($n_{\mathrm{rot,crit}}\simeq 10^{3}\ 
\mathrm{cm}^{-3}$). On the other hand, the vibrational levels are only
thermalized at densities greater than 10$^{10}$ cm$^{-3}$.  Therefore,
each vibrational level can be characterized by its own vibrational
temperature $T_{\mathrm{vib}}$. A large range of temperatures was
explored ($T$= 1000-4000~K) and the maximum temperature allowed for
the dust is set at 1500~K. The gas temperature was allowed to vary
with a power-law of index -3/4
($T(r)=T_{\mathrm{o}}(r/R_{\mathrm{min}})^{-3/4}$), typical for a flat
accretion disk \citep{Adams1988ApJ...326..865A} and the gas surface
density decreases with radius as $r^{-1}$
($N(r)=N_{\mathrm{o}}(r/R_{\mathrm{min}})^{-1}$) . The free parameters
of the models are the inclination $i$, the inner and outer radius
($R_{\mathrm{min}}$ and $R_{\mathrm{max}}$) of the hot CO emitting
area, and the temperature and CO column density at the inner radius
($T_{\mathrm{o}}$ and $N_{\mathrm{o}}$).  A grid of models was
performed and the model with the lowest reduced chi-square is shown in
the lower panel of Fig.~\ref{Fig_51Oph_fit}.  The gas is located in a
Keplerian disk seen with inclination angle of 88$^{+2}_{-35} \degr$
with respect to the rotational axis. The CO column density and
temperature at 0.15~AU are $(0.17-2.5)\ 10^{20}\ \mathrm{cm}^{-2}$ and
2850$\pm$500~K respectively.  The amount of CO gas can be inferred
from the modeling and is $M(\mathrm{CO}) \simeq (0.17-2.5)\ 10^{-10}\ 
M_{\odot}$. Our results show that the CO gas is located well within
the inner first AU from the central star
(Table~\ref{51Oph_table_fit}). The maximum distance $R_{\mathrm{max}}$
where the CO bandhead is emitted depends on the inclination, which is
not a well constrained parameter, but is probably smaller than the
dust sublimation radius $r_{\mathrm{d}}$ because the gas is at a much
higher temperature than the dust sublimation temperature.

\begin{table}[ht]
\begin{center}
\caption{Best fit parameters. $i$ is the inclination, $R_{\mathrm{min}}$ and $R_{\mathrm{max}}$ are the emission area inner and outer radius, $T_{\mathrm{o}}$(CO) is the gas temperature at $R_{\mathrm{min}}$ and $N_{\mathrm{o}}$(CO) is the column density of CO at $R_{\mathrm{min}}$.\label{51Oph_table_fit}}
\begin{tabular}{lllll}
\hline\hline
\noalign{\smallskip}
\multicolumn{1}{c}{$i$} & \multicolumn{1}{c}{$R_{\mathrm{min}}$} & \multicolumn{1}{c}{$R_{\mathrm{max}}$} & \multicolumn{1}{c}{$T_{\mathrm{o}}$(CO)} & \multicolumn{1}{c}{$N_{\mathrm{o}}$(CO)} \\
\multicolumn{1}{c}{($\degr$)} &\multicolumn{1}{c}{(AU)} &   \multicolumn{1}{c}{(AU)} & \multicolumn{1}{c}{(K)} & \multicolumn{1}{c}{(cm$^{-2}$)}  \\
\hline
\noalign{\smallskip}
88$^{+2}_{-35}$ & 0.15$\pm$0.05 & 0.35$\pm$0.05 & 2850$\pm$500 & $(0.17-2.5)\ 10^{20}$  \\
\noalign{\smallskip}
 \hline
 \end{tabular}  
\end{center}
\end{table}

\section{Discussion}\label{Discussion}

The observation of CO bandhead confirms the large column density of CO
seen by ISO-SWS. Moreover, the profile of the emission lines indicates
that the gas is seen {\em almost} edge-on. If the disk around
\object{51~Oph} is seen with an inclination of 88$^{+2}_{-35}$ and the
disk is geometrical thin (flat disk), then the ISO-SWS and the FUSE
data can be reconciled. In the FUSE absorption study, the
line-of-sight does not intercept the major part of the disk while
emission studies with ISO or ISAAC are not sensitive to the actual
projection angle $i$. On the other hand, hot gas of \ion{N}{i},
\ion{S}{ii} and \ion{Fe}{iii} can be detected in absorption against
the stellar photospheric emission because the evaporating bodies may
follow a trajectory that is inclined with respect to the equatorial
plane \citep{Beust2001A&A...366..945B}.

In the absence of extinction by dust grains, the principal destruction
agent of CO is photodissociation by stellar ultraviolet photons.
However, the CO column densities (10$^{20}$ -- 10$^{21}$ cm$^{-2}$)
are well above the required value for the CO molecules to self-shield
\citep[$N$(CO) $\sim
10^{15}$~cm$^{-2}$,][]{vanDishoeck1988ApJ...334..771V}.  Moreover, at
high-temperature and density the rapid formation of CO molecules by
the neutral-neutral reaction C+OH $\rightarrow$CO+H can easily
compensate for the photodestruction. The large abundances in H$_2$O,
CO$_2$ and CO testify of a hot and dense chemistry
\citep{Ancker2001A&A...369L..17V}.

It is interesting to compare the inner ($R_{\mathrm{min}}$) and outer
($R_{\mathrm{max}}$) CO emitting radius to other relevant radii in the
disk. The co-rotation radius $r_{\mathrm{rot}}=GM_*/\Omega_*^2$ is
defined as the distance from the central star where the star and the
Keplerian disk rotate at the same speed $\Omega_*$ (e.g.,
\citealt{Shu1994ApJ...429..781S}). Assuming that $\Omega_* = v \sin i
= 267$ km s$^{-1}$ (i.e. $i=90 \degr$), we obtain
$r_{\mathrm{rot}}\simeq 0.05$ AU (Table~\ref{table_comparison}). Disk
material falls on to the star surface only when the disk truncation,
whose value is close to the magnetospheric radius $r_{\mathrm{m}}$, is
smaller than the co-rotation radius for an accretion disk. The
magnetospheric radius depends on the stellar magnetic field and the
disk accretion rate, which are unknown for \object{51~Oph}. We
therefore assume that $r_{\mathrm{m}}\simeq r_{\mathrm{rot}}$. Dust
grains do not exist above their sublimation temperature, which is
around 1500~K. The dust sublimation radius is the distance from the
star beyond which dust can condense:
$r_{\mathrm{d}}=\sqrt{Q_R(L_*+L_{\mathrm{acc}})/16 \pi
  \sigma}/T_{\mathrm{sub}}^{2}$ where
$Q_R=Q_{\mathrm{abs}}(a,T_*)/Q_{\mathrm{abs}}(a,T_{\mathrm{sub}})$ the
ratio of the dust absorption efficiency at stellar temperature $T_*$
to its emission efficiency at the dust sublimation temperature
$T_{\mathrm{sub}}$, $a$ is the mean grain radius and $\sigma$ is the
Stefan constant. For large silicate grains ($a \geq 1\ \mu$m), $Q_R$
is relatively insensitive to the stellar effective temperature and
close to unity because most stellar radiation lies at wavelengths
shorter than the grain radius. For smaller grains, the value of $Q_R$
is significantly increased \citep{Monnier2002ApJ...579..694M}. We
found for \object{51~Oph} that $r_{\mathrm{d}}= 0.56$ AU (see
Table~\ref{table_comparison}) assuming $L_{\mathrm{acc}}<<L_* = 260 \ 
L_\odot$ and $Q_R \simeq 1$ (large grains). If the grains were as
small as 0.1 $\mu$m in radius, then $Q_R \simeq 20$ and
$r_{\mathrm{d}}= 4.50$~AU.

The disk around \object{51~Oph} is not strongly accreting as testified
by the absence of UV excess. The disk is passively heated with the gas
in the upper layer being warmer than the mid-plane. It should be
noticed that the dust sublimation radius is a lower limit where dust
grains can exist.  Radiative pressure, which is particularly effective
for a B9.5V star, will push the grains much further out than
$r_{\mathrm{d}}$.  At radii greater than the dust sublimation radius,
the dust and gas are probably thermally coupled and emission in the
near-infrared will be dominated by the dust, reducing drastically the
gas emission lines over continuum contrast. The large difference
between $r_{\mathrm{d}}$ and $r_{\mathrm{rot}}$ (0.5-2.45~AU) implies
the presence of a gas-rich dust-poor inner disk extending from the
truncation radius, which is close to the magnetospheric radius, till
the dust sublimation radius. It should be noticed that beyond 1~AU the
gas will become too cold to excite the high vibrational levels
($v>2$). On the other hand high-$J$ CO fundamental lines ($v=1-0$  
centered at $\lambda \sim$ 4.67~$\mu$m) are sensitive to gas at
temperature $\sim$1500~K.
\cite{Blake2004ApJ...606L..73B,Brittain2003ApJ...588..535B} found CO
gas at $T$ $\simeq$~1000--1500~K in the disk around \object{HD~163296}
and \object{AB~Aur}. Likewise, CO gas at $\sim$1500~K was detected by
\cite{Ancker2001A&A...369L..17V} toward \object{51~Oph}.
  
The opacity of a dust-poor gas at $T_{\mathrm{gas}}=2000-3000$~K is
dominated by H$_2$O and H$_2$, which may eventually deprive the inner
dust rim of large amount of stellar UV photons below 1215 \AA.
Indeed, the SED of \object{51~Oph} is well fitted by the emission from
an optically thick geometrically flat disk and does not show
significant near-infrared bump, consistent with the absence of a
puffed-up rim (\citealt{Dullemond2001ApJ...560..957D};
\citealt{Meeus2001A&A...365..476M}).
  
In summary, the presence of hot CO, traced by the bandhead emission,
and at the same time the absence of inner puffed-up rim are probably
linked, requiring that $r_{\mathrm{rot}}<<r_{\mathrm{d}}$.  This
criterion implies simultaneously a {\em high total luminosity}
$L=L_*+L_{\mathrm{acc}}$ \citep{Natta2001A&A...371..186N} and a
relatively {\em fast rotating} star. Interestingly, this criterion has
also been advocated to explain the general absence of inner puffed-up
rim in most T~Tauri disks \citep{Muzerolle2003ApJ...597L.149M}. We
compare the thickness of the dust free region of \object{51~Oph} with
two classical Herbig~Ae stars in Table~\ref{table_comparison} for
grains with radius of 0.1~$\mu$m and 1~$\mu$m.  It is clear that
\object{51~Oph} constitutes a special case among the Herbig~Ae stars
since it combines a large rotation velocity and a large luminosity.
Indeed CO bandhead emission has not been detected toward
\object{HD~163296} and \object{AB~Aur} (R.  Waters, private
communication). The anti-correlation between CO bandhead emission and
puffed-up rim should be tested in a larger sample of young stars
surrounded by a disk.
  
\begin{table}[ht]
\begin{center}
\caption{Comparison of the co-rotation and dust sublimation radius between \object{51~Oph} and two well studied Herbig~Ae stars. The spectral type and rotation velocity for \object{HD~163296} and \object{AB~Aur} are taken from \cite{Mora2001A&A...378..116M}. The other stellar characteristics are provided by \cite{Ancker1998A&A...330..145V,Ancker2001A&A...369L..17V}. The accretion luminosities were measured by \cite{Hillenbrand1992ApJ...397..613H}. The fits to the Spectral Energy Distribution of \object{HD~163296} and \object{AB~Aur} are discussed in \cite{Dominik2003A&A...398..607D}.\label{table_comparison}}
\begin{tabular}{llll}
\hline\hline
\noalign{\smallskip}
 &\multicolumn{1}{c}{\object{51~Oph}}&\multicolumn{1}{c}{\object{HD~163296}} & \multicolumn{1}{c}{\object{AB~Aur}}\\
\hline
\noalign{\smallskip}
SpT                                    & B9.5IIIe& A3Ve & A0Ve+sh\\
M$_*$ ($M_{\odot}$)                    & 3.8     & 2.0$\pm$0.2 & 2.4$\pm$0.2 \\
$L_*$  $(L_{\odot}$)                   & 260$^{+60}_{-50}$ & \phantom{1}26$\pm$5 &  47$\pm$12\\
$L_{\mathrm{acc}}$  $(L_{\odot}$)      & ... & \phantom{1}23 &  28 \\
$v\sin{i}$ km s$^{-1}$                 & 267$\pm$5  & 133$\pm$6& 97$\pm$20\\
$r_{\mathrm{rot}}$ (AU)                & 0.05 & 0.13 & \phantom{-}0.33 \\
$r_{\mathrm{d}}$ (AU)~(a=1.0~$\mu$m)& 0.56 & 0.18 & \phantom{-}0.24 \\
$r_{\mathrm{d}}$ (AU)~(a=0.1~$\mu$m)& 2.50 & 0.80 & \phantom{-}1.07 \\
$r_{\mathrm{d}}-r_{\mathrm{rot}}$ (AU)~(a=1.0~$\mu$m)  & 0.51 & 0.05 & -0.09 \\
$r_{\mathrm{d}}-r_{\mathrm{rot}}$ (AU)~(a=0.1~$\mu$m)  & 2.45 & 0.38 & \phantom{-}0.74 \\
CO bandhead emission                   & yes & no & no\\
CO fundamental emission                   & yes & yes & yes\\
Puffed-up inner rim                    & no & yes & yes \\
\noalign{\smallskip}
\hline
\end{tabular}  
\end{center}
\end{table}
%

\section{Conclusions}

The principle results are summarized here:
\begin{itemize}
\item A high column density of hot molecular (CO) gas is present in
  the inner Astronomical Unit of \object{51 Oph}. The gas rotates
  around the star in a Keplerian orbit.
\item The apparent discrepancy on the amount of CO gas around
  \object{51 Oph} between the ISO-SWS and FUSE data can be resolved if
  the disk is geometrically flat and is seen with a large angle but
  not entirely edge-on.
\item The CO bandhead emission is located inside the dust sublimation
  radius. We propose that only star-disk systems where the co-rotation
  radius is much smaller than the dust sublimation radius
  ($r_{\mathrm{rot}}<<r_{\mathrm{d}}$) will show high line over
  continuum contrast CO bandhead emission. This later prediction can
  be tested with future high resolving power near-infrared
  interferometric instruments like Amber at the VLT. \object{51 Oph} is
  a special case since its rotation speed of 267 km s$^{-1}$ is much
  higher than the average value for Herbig~Ae stars (80--150 km
  s$^{-1}$).
\item Another test for the existence of the dust-free gas would be the
  detection of strong H$_2$ fluorescence lines in the UV (e.g.,
  \cite{Herczeg2004ApJ...607..369H} who found H$_2$ UV lines in the
  inner disk around the T~Tauri star \object{TW~Hya}). The observation
  of near-infrared H$_2$ lines with the $v$=2-1 S(1) intensity at
  2.2477 $\mu$m higher than the $v$=1-0 S(1) at 2.1218 $\mu$m would
  also provide another diagnostic of dust-free gas.
\end{itemize}

\begin{acknowledgements} 
  WFT is supported by NWO grant 614.041.005.  The authors thank the
  VLT staff for performing the observations in Service mode.
\end{acknowledgements}


\bibliographystyle{aa}
\bibliography{51Oph}

\end{document}